\documentclass[twocolumn,showpacs,amsmath,amssymb]{revtex4}
\usepackage{graphicx}
\usepackage{dcolumn}
\usepackage{bm}
\usepackage{amsmath}
\usepackage{amsfonts}
\begin{document}

\title{Composite Millicharged Dark Matter}
\author{Chris {\sc Kouvaris}}\email{kouvaris@cp3.sdu.dk}
\affiliation{$\text{CP}^3$-Origins, University of Southern Denmark, Campusvej 55, Odense 5230, Denmark}

\begin{abstract}
We study a composite millicharged dark matter model. The dark matter is in the form of pion-like objects emerging from a higher scale QCD-like theory. We present two distinct possibilities with interesting phenomenological consequences based on the choice of the parameters. In the first one, the dark matter is produced non-thermally and it could potentially account for the 130 GeV Fermi photon line via decays of the ``dark pions''. We estimate the self-interaction cross section which might play an important role both in changing the dark matter halo profile at the center of the galaxy and in making the dark matter warmer. In the second version the dark matter is produced via the freeze-in mechanism. Finally we impose all possible astrophysical, cosmological and experimental constraints. We study in detail generic constraints on millicharged dark matter that can arise from  anomalous isotope searches of different elements and we show why constraints based on direct searches from underground detectors are not generally valid. 
\\[.1cm]
 {\footnotesize  \it Preprint: CP$^3$-Origins-2013-15 DNRF90 \& DIAS-2013-15.}\end{abstract}


\maketitle 

\section{Introduction}

The current knowledge is that all particles in the Standard Model are either electrically neutral (e.g. neutrinos, photons and gluons), or have charges which are integer multiples of $e/3$ where $e$ is the charge of the electron. The possibility of particles with electric charges much smaller than $e/3$ has always intrigued both theorists and experimentalists. One way one can add millicharged particles in the Standard Model is for example by adding a particle with hypercharge $Y=2\epsilon$ that is singlet of the color $SU(3)$ and weak $SU(2)$. However as it was pointed out in~\cite{Okun:1983vw}, the embedment of such an extension of the Standard Model into a Grand Unified Theory is problematic. These problems can be avoided if an extra $U(1)'$ is introduced~\cite{Holdom:1985ag}. Due to mixing between photons and the paraphotons of  the $U(1)'$, particles charged under the $U(1)'$ appear to have a small coupling to the photon and thus as electrically millicharged. A third way millicharged particles might appear is in the form of Standard Model neutrinos that have tiny electric charges. A small modification of the hypercharge assignments of the Standard Model particles can accommodate tiny electric charges to neutrinos without inducing any gauge anomaly (for a review see e.g. \cite{Foot:1989fh,Foot:1992ui,Davidson:2000hf}).

 Although Charged Massive Particles (ChaMPs) have been more or less ruled out as dark matter candidates, this is not easily the case for particles with tiny charges. Millicharged particles have been firstly proposed  as a solution of the dark matter puzzle long time ago~\cite{Goldberg:1986nk,DeRujula:1989fe,Dimopoulos:1989hk}. Composite dark matter (with charged constituents) is an attractive scenario that has been studied extensively~\cite{Nussinov:1985xr,Barr:1990ca,Gudnason:2006yj,Gudnason:2006ug,Kouvaris:2007iq,Kainulainen:2009rb}. 
 Recently there has been a revived interest on millicharged dark matter via Stueckelberg $Z'$ models~\cite{Cheung:2007ut,Feldman:2007wj}, in the form of millicharged atomic dark matter~\cite{Cline:2012is}, or in the form of MeV particles in order to explain the 511 keV line~\cite{Huh:2007zw}.

In this paper we present a model of composite dark matter inspired by QCD. As in QCD where the spontaneous symmetry breaking of the chiral symmetry leads to Goldstone bosons, similarly here, chiral symmetry breaking produces Goldstone bosons that play the role of dark matter. We explore the phenomenology of this theory examining two interesting cases. In the first one, the ``dark pions'' are produced non-thermally and they can explain the 130 GeV Fermi photon line via decays of the form $\pi_0\rightarrow 2 \gamma$. In the second version, we show how the pions can be produced thermally via the freeze-in mechanism. We also study all possible astrophysical, experimental and cosmological constraints imposed on such a model. 

We chose a QCD-type Lagrangian of the form
\begin{equation}
L=\bar{\psi}i\gamma_{\mu}D^{\mu}\psi - m\bar{\psi}\psi -\frac{1}{4}F^{\alpha \mu \nu}F^{\alpha}_{\mu \nu},
\end{equation} 
where the gauge group is an $SU(3)$ ``color''. We assume two flavors of fermions that carry tiny electric charges  $2\epsilon e/3$ and $-\epsilon e/3$ where $\epsilon$ is a small number that will be estimated later on. The covariant derivative is
$D^{\mu} = \partial^{\mu} -ig \tau^{\alpha} G^{\alpha \mu} -i q A^{\mu}$, where $G^{\alpha \mu}$ is the ``new gluon'' field and $A^{\mu}$ is the electric field.  We assume that the theory becomes confined below a scale, where the description of the theory in terms of mesonic and baryonic degrees of freedom is more appropriate. There are three Goldstone bosons i.e. the pions of the theory $\pi^+$, $\pi^-$, and $\pi_0$. As in QCD, it is  expected (due to electromagnetic interactions) that  $\pi^{\pm}$ are heavier than $\pi_0$. 
For simplicity we are going to assume that there are no baryons of the theory present, and therefore dark matter is composed only of the three pions. This is expected in cases where the ``new baryon'' number is not conserved.  We study now the phenomenological implications of this simple QCD inspired dark matter model, starting from the possibility of producing the Fermi 130 GeV line.

\section{The 130 GeV line}
Recently, analysis of the Fermi collaboration data revealed the possibility of the existence of a $\gamma$ line at 130 GeV coming from the center of the galaxy~\cite{Weniger:2012tx,Bringmann:2012vr,Tempel:2012ey,Boyarsky:2012ca,Su:2012ft,Hektor:2012kc}. Due to the fact that this line is sufficiently peaked and 
does not follow the typical astrophysical background power law behavior, this can be an indirect signature of dark matter presence. In general,
 if dark matter is in the form of WIMPs (denoted by $\chi$), processes like $\chi+ \chi \rightarrow \gamma+\gamma$, $\chi+ \chi \rightarrow \gamma+X$, or $\chi \rightarrow \gamma+\gamma$ are in principle all allowed. Obviously, in the case of annihilating dark matter to two photons, the WIMP mass must be 130 GeV, and in the decaying dark matter scenario, the mass must be 260 GeV. For the case where one photon and one other particle $X$ is produced, the WIMP mass depends obviously on the mass of $X$.   One striking observation is that if the line is
 attributed to annihilation of WIMPs, the annihilation cross section does not seem to be the one required for thermal production of WIMPs, but it is rather smaller by roughly an 
 order of magnitude depending on the assumed dark matter halo profile. This could be an indication towards an asymmetric WIMP annihilation where dark matter and anti-dark matter do not appear in nature with equal numbers (something supported by some recent analysis~\cite{Masina:2013yea}) or towards a decaying dark matter scenario. However, in both cases a large branching ratio to photons (larger than $\sim 0.01$) is required~\cite{Buchmuller:2012rc,Cohen:2012me,Cholis:2012fb}. Otherwise, decays and/or annihilations to other particles can induce a continuum spectrum of photons that can hide the line. From this perspective a dark pion that has only the decay channel $\pi_0 \rightarrow 2 \gamma$ fulfills perfectly this requirement.

 Although Einasto or Navarro-Frenk-White dark matter halo profiles favor the scenario of annihilating dark matter~\cite{Weniger:2012tx}, decaying dark matter could account for the 130 GeV line if the dark matter morphology at the center of the galaxy is appropriate (i.e. more cuspy). The photon spectrum of a decaying WIMP to two photons is

\begin{equation}
\frac{d\Phi_{\text{dec}}}{dE_{\gamma}d\Omega}=\frac{\Gamma}{4\pi}r_{\odot} \left (\frac{\rho_{\odot}}{2m_{\chi}} \right ) \int_{\text{los}}ds\frac{1}{r_{\odot}} \left ( \frac{\rho_{\text{halo}}(r)}
{\rho_{\odot}} \right ) \frac{dN_{\text{dec}}}{dE_{\gamma}},
\end{equation}
while for the annihilating one

\begin{equation}
\frac{d\Phi_{\text{ann}}}{dE_{\gamma}d\Omega}=\frac{\langle \sigma v \rangle}{8\pi}r_{\odot} \left (\frac{\rho_{\odot}}{m_{\chi}} \right )^2 \int_{\text{los}}ds\frac{1}{r_{\odot}} \left ( \frac{\rho_{\text{halo}}(r)}
{\rho_{\odot}} \right )^2 \frac{dN_{\text{ann}}}{dE_{\gamma}},\end{equation}
where $\Gamma$, and $\langle \sigma v \rangle$ are the decay rate and 
annihilation cross section to two photons respectively, $dN/dE_{\gamma}$ denotes differential photon energy spectrum, $\rho_{\odot} \simeq0.3~\text{GeV}/\text{cm}^3$ is the local dark matter density, $r_{\odot}\simeq8.5~\text{kpc} $ is our distance from the center of the galaxy and $\rho_{\text{halo}}(r)$ is the dark matter halo profile. The integral is taken along the line of sight. 
Decaying dark matter has already been proposed as an explanation of the 130 GeV line~\cite{Kyae:2012vi,Kang:2012bq,Park:2012xq,Fan:2012gr}.

In this paper we do not attempt to make a sophisticated analysis of the Fermi spectrum. Following~\cite{Kyae:2012vi} on generic grounds, we assume that the observed spectrum is attributed to decaying dark matter if one adjusts the decaying rate such that the spectrum will be similar to the annihilating one. Depending on the chosen halo profile, an approximate value of $\Gamma_{\chi \rightarrow \gamma \gamma}=10^{-29}\text{sec}^{-1}$ can satisfy the requirement~\cite{Kyae:2012vi}. We should mention that the analysis in~\cite{Weniger:2012tx} suggests that annihilating dark matter fits much better the Fermi spectrum compared to a decaying dark matter scenario. However a decaying dark matter scenario can be valid  if the dark matter density is more cuspy at the galactic center. Although this is not currently favored by numerical simulations of collisionless dark matter, the existence of self-interactions might alter completely the situation. We shall show that indeed such self-interactions are present among the dark pions.  

The $\pi_0$ of the theory decays due to the anomaly exactly as in QCD. However, due to the fact that the constituent fermions have tiny charges, the lifetime time of the particle can change significantly. The decay rate is
\begin{equation}
\Gamma_{\pi_0 \rightarrow \gamma \gamma}=\frac{\alpha^2 \epsilon^4}{64\pi^3} \frac{m_{\pi_0}^3}{f_{\pi}^2},
\end{equation}
where $\alpha$ is the usual fine structure constant. Demanding the decay rate to be $\sim 10^{-29}~\text{sec}^{-1}$ gives the condition
\begin{equation}
\frac{\epsilon}{\sqrt{f_{\pi}}} \simeq 6 \times 10^{-14}, \label{eps2}
\end{equation}
where $f_{\pi}$ is measured in GeV. For a typical value of $f_{\pi}=1$ TeV, $\epsilon \simeq 1.9 \times 10^{-12}$. 
The pions can also co-annihilate. In order to find what is the annihilation cross section, we should introduce an effective chiral perturbation Lagrangian that describes the mesonic fields. Similarly to QCD
\begin{equation}
L= \frac{f_{\pi}^2}{4}\text{Tr}(D_{\mu}\Sigma D^{\mu} \Sigma^{\dagger})+ \frac{f_{\pi}^2}{2}\text{Tr}(m_{\pi}^2 \Sigma ), \label{ChPT}
\end{equation}
where $D_{\mu}=\partial_{\mu}+ieA_{\mu}[Q, \Sigma]$, $Q$ being a diagonal matrix with entries the electric charges of the two ``quarks'' in units of $e$. The matrix $\Sigma$ is defined as usual
\begin{equation}
\Sigma= \exp \left ( i\frac{\tau^\alpha \pi^\alpha}{f_{\pi}} \right ),
\end{equation}
where $\tau^\alpha$ are the Pauli matrices. It is easy to see that to lowest order $\pi_0$ does not couple to photons (as it is expected), because the commutator $[Q, \Sigma]$ is zero.
However, $\pi_0$'s can annihilate to photons through one-loop mediation of $\pi^\pm$ states that couple to photons. There are two Feynman diagrams contributing to the amplitude. 
For on-shell production of photons the total amplitude is
\begin{equation}
M=M_{\mu \nu} \epsilon^{1\mu} \epsilon^{2\nu},
\end{equation} where $\epsilon^{1,2}$ represent the helicity of the two produced photons and $M_{\mu \nu}$ is~\cite{Donoghue:1988eea}
\begin{eqnarray}
M_{\mu \nu}  =  \frac{-i}{16 \pi^2}  \frac{2e^2 \epsilon^2}{f_\pi^2} \frac{s-m_\pi^2}{s} (g_{\mu \nu} s-2 k_{2 \mu} k_{1 \nu} ) \nonumber \\
 \times \left \{1+ \frac{m_\pi^2}{s} \left [\text{ln} \left [ \frac{z_+}{z_-} \right ] -i \pi \right ]^2 \right \}, \label{magn}
 \end{eqnarray} where $z_{\pm}=1\pm (1-4m_\pi^2/s)^{1/2}$, $k$ represents photon momentum, and $s$ is the typical Mandelstam variable. It is understood that for the derivation of this result, we have made the approximation that the mass difference between $\pi^{\pm}$ and $\pi_0$ is not large.   Given the above amplitude for $\pi_0 \pi_0 \rightarrow \gamma \gamma$, we can compute the corresponding cross section in the center of mass frame 
 \begin{equation}
   \langle \sigma v \rangle =\frac{\alpha^2 \epsilon^4}{32 \pi^3 f_\pi^4} \frac{(s-m_\pi^2 )^2}{s}\left [ 1+\frac{m_\pi^2}{s}f(s) \right ],
 \end{equation} 
 where $f(s)=2[\text{ln}^2(z_+/z_-)-\pi^2]+(m_{\pi}^2/s)[\text{ln}^2(z_+/z_-)+\pi^2]^2$.
   Upon making the assumption that $\pi_0$ move with non-relativistic velocities ($s\simeq 4m_{\pi}^2$ and $z_{\pm}\simeq 1$), and by using Eq.~(\ref{eps2}), 
   \begin{equation}
   \langle \sigma v \rangle \simeq 8.9 \times 10^{-89} \left (\frac{\text{TeV}}{f_\pi} \right )^2~\text{cm}^2.
   \end{equation} This annihilation cross section which in principle can produce a photon line at 260 GeV is so tiny to be detected or to play any role in the evolution of the dark matter density. 
    
  It is easy to show that $\pi$'s are not in thermal equilibrium with the plasma for the $\epsilon$ values we are interested. We can have a rough estimate of the Thomson cross section between a charged pion and a photon. The condition for thermal equilibrium is that $n_{\gamma} \sigma_T v  \gtrsim H$, where $n_{\gamma}$ is the number density of photons. The 
Thomson cross section (in natural units) is $\sigma_T=(8\pi/3m^2)\alpha^2 \epsilon^4$. Using $n_{\gamma}= (2\zeta (3)/\pi^2)T^3$ and $H=1.66 g_*^{1/2}T^2/m_{\text{pl}}$, it is easy to show that the dark pions (or dark fermions before confinement) with $\epsilon < 10^{-5}$ were never in thermal equilibrium  with the plasma (unless one goes to extremely high temperature when pions did not exist). This means that in this case $\pi$'s have to be produced non-thermally by some unknown mechanism. Whatever the mechanism is, it should not disturb the thermal evolution of the universe (i.e. the effective degrees of freedom $g_*$ should not change). This can easily happen for example if the pions are produced via some decay with relativistic energies but with a temperature significantly lower than that of the plasma. As we pointed out above, the plasma is not in thermal equilibrium with the pions and therefore they do not have to have the same temperature. However pion self-interactions are sufficient to establish thermal equilibrium among the pions providing pions with a common temperature that must be much smaller than the temperature of the plasma in order to avoid problems with the evolution of the universe. The reason that self-interactions can keep pions in thermal equilibrium is due to the fact that the cross section is independent of $\epsilon$ and therefore is not suppressed. 
The coupling among pions can be estimated from Eq.~(\ref{ChPT}) if one expands the second term to fourth order in the fields. The amplitude for $\pi_0 \pi_0 \leftrightarrow \pi_0 \pi_0$ (but also  for $\pi^+ \pi^- \rightarrow \pi_0 \pi_0$ if one assumes that the difference in the masses of $\pi_0$ and $\pi^+$ is small) is
\begin{equation}
M= \frac{1}{f_\pi^2}  (s-m_{\pi_0}^2). \label{amp_pipi}
\end{equation}
This leads to a cross section
\begin{equation}
\langle \sigma v \rangle = \frac{1}{16§ \pi f_\pi^4}\frac{(s-m_{\pi_0}^2)^2}{s}\sqrt{1- \frac{4m_{\pi_0}^2}{s}}. \label{pipi_cross}
\end{equation}
The condition $n_{\pi} \sigma_{\pi\pi} v>> H$ is easily fulfilled. One can check by using the number density for a relativistic species with a temperature $T_{\text{dm}}<<T_{\gamma}$ and the above cross section, that pions were in thermal equilibrium with each other at some point. Obviously as the universe expands and the pions become non-relativistic, pions at some point will lose contact with each other, but as in the case with neutrinos, they will preserve their Bose-Einstein distribution since their entropy is preserved. The fact that pions were in thermal equilibrium enable us to estimate how much $\pi^{\pm}$ is left today. Assuming that    
$\pi^{\pm}$ are slightly heavier than $\pi_0$ (as it happens in QCD), the annihilation cross section for $\pi^+\pi^-\rightarrow \pi_0\pi_0$ is given again by Eq.~(\ref{pipi_cross}). The difference between the annihilation and the self-interactions among pions is that the square root in Eq.~(\ref{pipi_cross}) is $\sim \sqrt{ 1-m_{\pi_0}^2/ {m_\pi^{\pm 2}}}$ for the former and $\sim v$ for the latter. It is understood that in the case of annihilation at the nonrelativistc limit $s\simeq 4 m_{\pi^{\pm}}^2$. In the nonrelativistic limit, the annihilation cross section takes the form
\begin{equation}
\langle \sigma v\rangle \simeq 2.3 \times 10^{-36}\text{cm}^2 \frac{m_{270}^2}{f_1^4}\sqrt{1-0.927/m_{270}^2}, \label{pi0_prod}
\end{equation} 
where $f_1$ is the value of $f_{\pi}$ in units of 1 TeV, and $m_{270}$ is $m_{\pi^{\pm}}$ in units of 270 GeV. This is quite an interesting result. By playing with $f_{\pi}$ and the mass splitting between $\pi^{\pm}$ and $\pi_0$, one can get from complete extinction of $\pi^{\pm}$ today to a partial one. In the case of partial presence of $\pi^{\pm}$, assuming that $x$ fraction of dark matter is in the form of $\pi_0$ and $1-x$ in the form of $\pi^{\pm}$, Eq.~(\ref{eps2}) should be modified to
\begin{equation}
\frac{\epsilon}{\sqrt{f_{\pi}}} \simeq 6 \times 10^{-14}\frac{1}{x^{1/4}}.
\end{equation}
This is because the Fermi 130 GeV line has to come from decay of the $x$ part of the dark matter density (i.e. $\pi_0$). There are two comments in order here. Due to the smallness of $\epsilon$, no essential constraints can be imposed on the $1-x$ part of dark matter that is made of $\pi^{\pm}$ (see the constraint section).

 The second comment has to do with the fact that the freeze-out of the annihilation takes place at a temperature $T_{\text{dm}}$ which as we shall show below must be significantly smaller than the plasma temperature $T_{\gamma}$. However, since everything takes place in high enough energies, $g_*$ and $g_{*s}$ do not change substantially, and therefore one can approximately consider that the freeze-out occurs  
 as usually at $T_{\text{dm}}\sim m_{\pi^{\pm}}/25$. However one should keep in mind that this corresponds to a much higher plasma temperature. One would naively expect that earlier freeze-out would correspond to even colder dark matter. This is not necessarily true due to self-interactions. We discuss this issue later in this section.
 
As we mentioned $T_{\text{dm}}$ is in principle different than  $T_{\gamma}$. If pions have been produced relativistically, one has to make sure that by the time of the Big Bang Nucleosynthesis (BBN), dark matter does not change significantly the effective number of relativistic degrees of freedom $g_*$. In order to find how the constraint affects our dark matter scenario, let us assume that the pions are produced by a time when their temperature was $T_{d1}$ and the plasma temperature was $T_1$. Let us call $\beta=T_{d1}/T_1$. Now let's assume that the dark pions become nonrelativistic at some time where their temperature is $T_{d2}\simeq m$ and the plasma's temperature is $T_2$. Up to that point both the plasma and the pions are relativistic and therefore the ratio $T_{d2}/T_2$ is still $\beta$. Below that temperature pions become non-relativistic. However,  since pions were never in thermal equilibrium with the plasma, their entropy should be conserved. At the BBN time scale, entropy conservation dictates $n_{\text{BBN}}a_{\text{BBN}}^3=n_2a_2^3$ (where 2 refers to the time where pions become nonrelativistic). The energy density of pions at the BBN time scale is $\rho_{\text{BBN}}=mn_{\text{BBN}}=n_2m(a_2/a_{\text{BBN}})^3=n_2m(T_{\text{BBN}}/T_2)^3$, where $T_{\text{BBN}}=1$ MeV. Using the fact that $n_2=\zeta(3)T_{d2}^3/\pi^2$ and that $T_{d2}/T_2=\beta,$ one can derive the contribution of pions to $g_*$ at BBN as
\begin{equation}
g_{* \text{dm}}^{\text{BBN}}=\frac{30}{\pi^4}\zeta(3)\beta^3\frac{m}{T_{\text{BBN}}}.
\end{equation} 
Extra relativistic degrees of freedom at the epoch of BBN are tightly constrained because the abundances of light elements are strongly affected by $g_*$. At BBN $g_*=5.5 + 7/4*N_\nu$ where $N_\nu$ is the number of neutrino families.  Analysis of the BBN light element abundances and CMB data from Planck limit $N_\nu$ down to $\sim 0.5$~\cite{Ade:2013lta} (on top of the three existing neutrino families). This leads to a freedom of $g_{* \text{dm}}^{\text{BBN}}<(7/4)0.5$. This  dictates that $\beta< 0.02$ in order not to destroy the BBN predictions. We can now estimate what is the $\beta$ needed in order to get the observed relic density of dark matter today. In order to do this, we use again the conservation of the dark matter entropy $n_{\text{BBN}}a_{\text{BBN}}^3=n_0$, where $n_0=1.176 \times 10^{-6}\text{GeV}/\text{cm}^3$ is the dark matter density today. This leads to
\begin{equation}
n_0=\frac{\zeta(3)}{\pi^2}\beta^3 T_{\text{BBN}}^3a_{\text{BBN}}^3. \label{n0}
\end{equation}
The only unknown $a_{\text{BBN}}$ can be easily estimated by using conservation of the entropy for the plasma this time i.e. $g_{*s}^{\text{BBN}}T_{\text{BBN}}^3a_{\text{BBN}}^3=g_{*s}^0T_0^3$. Using $g_{*s}^{\text{BBN}}=10.75$, $g_{*s}^0=3.909$ and $T_0=2.728$ K, we can get $a_{\text{BBN}}=1.68 \times 10^{-10}$. Using Eq.~(\ref{n0}) leads to a value $\beta \simeq 4 \times 10^{-4}$ which is much smaller than the 0.02 constraint we derived above.

As we have already mentioned, self-interactions are independent of $\epsilon$ and are not suppressed by powers of it. This can be quite crucial for two reasons: Firstly self-interactions might change the dark matter halo profile at the center of the galaxy. Although with standard dark matter halo profiles, Fermi data~\cite{Weniger:2012tx} favor annihilating dark matter over decaying,  dark matter self-interactions might change completely this picture, since most dark matter simulations so far consider collisionless dark matter. The second point is that self-interactions might make the dark pion to be a warmer  dark matter candidate which is supported by some data~\cite{Lovell:2011rd}. The nonrelativistic self-interaction cross section can be read off from Eq.~(\ref{pipi_cross})
\begin{equation}
\sigma_{\text{self}}\simeq 5.9 \times 10^{-37} \left (\frac{1\text{TeV}}{f_\pi} \right )^4 \left (\frac{m_{\pi_0}}{260\text{GeV}} \right )^2 \text{cm}^2.
\end{equation}
The value is sufficiently small to enable the dark pion to evade constraints on dark matter self-interactions imposed by the bullet cluster~\cite{Randall:2007ph}, or galaxy ellipticity~\cite{Feng:2009mn,Feng:2009hw}, or old neutron stars~\cite{Kouvaris:2011fi,Kouvaris:2011gb,Bell:2013xk}. However, the cross section being of the order of picopbarn or even higher (if one selects a smaller $f_\pi$ and/or larger $m_{\pi_0}$), can potentially have an effect on how warm the dark matter candidate is, and on the inner dark matter halo profile. The answer to especially the second issue is difficult and in principle requires simulations of self-interacting dark matter which might be considerable different from the collisionless case. 

\section{Thermally produced millicharged Dark Matter}

Despite the tiny electric charges of its constituents, and the fact that $\pi$'s are not in thermal equilibrium with the plasma for small values of $\epsilon$, they can be produced thermally via the freeze-in mechanism~\cite{Hall:2009bx}
. The process $\gamma \gamma \rightarrow \pi^+ \pi^-$ is sufficient to produce an abundance similar to the dark matter one. The Boltzmann equation for the process above  can be written as~\cite{Hall:2009bx}
\begin{widetext}
\begin{equation}
\frac{dn_{\pi^{\pm}}}{dt}+3Hn_{\pi^{\pm}}=\frac{2T}{512\pi^6}\int^{\infty}_{4m_{\pi^{\pm}}^2}ds d \Omega P_{\gamma \gamma}P_{\pi^+ \pi^-} \lvert M \rvert^2_{\gamma \gamma \rightarrow \pi^+ \pi^-}K_1(\sqrt{s}/T)/\sqrt{s}, \label{freeze-in}
\end{equation}
\end{widetext}
where \begin{equation}
P_{ij}=\frac{[s-(m_i+m_j)^2]^{1/2}[s-(m_i-m_j)^2]^{1/2}}{2\sqrt{s}}.
\end{equation}
It is not hard to show that the leading $s$ contribution of the cross section comes from the Feynman diagram with the quartic coupling $\epsilon^2 e^2 A_{\mu}A^{\mu}\pi^+ \pi^-$. This diagram contributes to $\lvert M \rvert^2_{\gamma \gamma \rightarrow \pi^+ \pi^-}=2e^4 \epsilon^4$.   
Integrating Eq.~(\ref{freeze-in}) gives
 \begin{equation}
\frac{dn_{\pi^{\pm}}}{dt}+3Hn_{\pi^{\pm}}=s_1\frac{dY}{dt}=\frac{T^2m_{\pi^{\pm}}^2e^4\epsilon^4}{64\pi^5}K_1(x)^2, \label{k1}
\end{equation}
where $x=m_{\pi^{\pm}}/T$, and $s_1=(2\pi^2/45)g_{*s}T^3$ is the total entropy. Using $t=0.301g_*^{-1/2}(m_{\text{pl}}/m_{\pi^{\pm}}^2)x^2$, the equation can be rewritten as
\begin{equation}
\frac{dY}{dx}=\frac{45\times 0.301 e^4 \epsilon^4}{64 \pi^7g_{*s}g_*^{1/2}}\frac{m_{\text{pl}}}{m_{\pi^{\pm}}} x^2K_1(x)^2.
\end{equation}
We can now integrate the equation from $x=0$ to infinity. Using 
\begin{equation}
\Omega_{\text{dm}} h^2=\frac{2mY_{\text{dm}}}{3.6 \times 10^{-9} \text{GeV}}\simeq 0.112,
\end{equation}
 and $g_* \simeq g_{*s}\simeq 100$, we get $\epsilon=1.3 \times 10^{-5}$. This means that with $\epsilon \sim 10^{-5}$, the abundance of $\pi^{\pm}$ is that of the dark matter. 
 From Eq.~(\ref{eps2}), one can realize that millicharged  dark matter could in principle be produced via the freeze-in mechanism and provide via decays a $\sim 130$ GeV line for $f_{\pi}\simeq  4.8 \times 10^{16}$ GeV which is the GUT scale. However, this is not possible.  By a simple rescaling of Eq.~(\ref{pi0_prod}), one can see that the value of $f_\pi$ is so large (corresponding to a tiny $\pi^+\pi^-\rightarrow \pi_0  \pi_0$ cross section) that makes impossible to create substantial quantities of $\pi_0$. In addition, we found that the direct production of $\pi_0$ from the process $\gamma \gamma \rightarrow \pi_0\pi_0$ is suppressed parametrically by $(m_{\pi_0}/f_{\pi})^4$. This can be seen by using Eq.~(\ref{magn}) and follow the steps leading to Eq.~(\ref{k1}) keeping only the leading term in $s$. This means again that no $\pi_0$ are produced and therefore the Fermi 130 GeV line cannot be explained by this version of millicharged dark matter. However, as we demonstrate in the next section, this thermally produced (via the freeze-in mechanism) millicharged dark matter evades all possible constraints arising from dark matter direct searches in underground detectors.

\section{Constraints on Millicharged Dark Matter}

The presence even of tiny electric charges on particles (and in particular dark matter ones) in nature can lead to severe constraints. This has been addressed long time ago in the context of ChaMPs but it was extended in the case of fractionally charged particles~\cite{Goldberg:1986nk,DeRujula:1989fe,Dimopoulos:1989hk,Davidson:1993sj,Davidson:2000hf,c1,c2}. These bounds have been obtained by a variety of different methods ranging from accelerator searches and dark matter searches to constraints from the Lamb shift, Big Bang Nucleosynthesis, the Supernova 1987A, plasmon decay in red giants and white dwarfs. In addition there are cosmological bounds based on the fact that $\Omega<1$. A summary of these bounds can be seen for example in~\cite{Davidson:2000hf}. All these bounds are avoided by both versions of the dark pion we analyze in this paper. In the 260 GeV decaying pion, $\epsilon$ is too small to be constrained by these bounds. The same is true for the thermally produced dark pions
which although have $\epsilon \sim 10^{-5}$, it is still too small to be excluded (especially if pions are embedded in a model with paraphotons).

In addition, there are severe constraints on $\epsilon$ based on the bullet cluster~\cite{Randall:2007ph}
and the ellipticity of galaxies like NGC 720~\cite{Feng:2009mn,Feng:2009hw}. The bullet cluster imposes an upper bound on dark matter self-interactions and since charged particles with charge $\epsilon e$ can exchange photons, an upper bound is set for $\epsilon$. Coulomb-type
self-interactions can also make the distribution of dark matter more isotropic and this leads to more spherical galaxies. The ellipticity of certain galaxies like NGC 720 imposes again an upper bound on $\epsilon$. Furthermore, millicharged particles can in fact cause changes in the Cosmic Microwave 
Background spectrum leading to very strict constraints on $\epsilon$~\cite{Dubovsky:2003yn,Burrage:2009yz}. Among the above three constraints, the CMB ones are the most strict. However, a millicharged particle with mass larger than $\sim 100$ GeV and $\epsilon \lesssim 10^{-5}$ clearly avoids it.
Below we will study in more detail two other sets of constraints i.e. the ones from direct search detection in underground detectors and the ones from searches of anomalous isotopes of different elements.

\subsection{Direct Detection}
Direct detection search experiments like Xenon~\cite{Aprile:2011hi} and CDMS~\cite{Ahmed:2009zw} have imposed severe constraints on the cross section between dark matter particles and nuclei of detectors that have been planted in deep sites in order to be shielded from unnecessary background radiation and thus false signals. Millicharged dark matter can exchange photons with the nuclei and therefore constraints on $\epsilon$ can arise. Such constraints have been studied in the context of mirror dark matter~\cite{Foot:2010hu} and in general~\cite{McDermott:2010pa}. In principle such a constraint might be quite severe. Millicharged particles scatter off charged nuclei with a Rutherford cross section. Of course sufficiently small values of $\epsilon$ lead to little scattering and thus particles will avoid detection. Based on the CDMS data, the analysis of~\cite{McDermott:2010pa} provided an upper bound $\epsilon \simeq 10^{-8}$ for millicharged dark matter with mass above 10 GeV, provided that millicharged particles are not evacuated from the galactic disc. This limit is of course evaded by our decaying dark pion with an $\epsilon \sim 10^{-12}$, but it could exclude the thermally produced dark pion that requires $\epsilon \sim  10^{-5}$. However, we show that the CDMS excluded $\epsilon$-region of~\cite{McDermott:2010pa} is not generally valid. We shall demonstrate that for intermediate values of $\epsilon$ (small but not too small), millicharged particles could lose most of their kinetic energy through electromagnetic interactions with atoms before reaching the detectors while traveling underground,  leading to Rutherford scattering with recoil energies way below the threshold of the detectors thus invalidating this type of constraints.

A non-relativistic millicharged particle can interact and lose energy via collisions with electrons according to the Bohr-Bethe-Bloch formula
 \begin{equation}
 \frac{dE}{dx}=-4 \pi n \frac{\epsilon^2 e^4}{m_e v^2}\ln \frac{2m_e v^2}{I}, \label{dEdx1}
 \end{equation}
 where $n=N_AZ\rho/M_u$ is the electron density of the target ($N_A$, $Z$, and $M_u$ being respectively the Avogadro number, the atomic number of the material, and the molar mass constant),  $m_e$ is the mass of the electron, and $I\simeq 10Z\rm eV$ is the mean ionization potential. However, at low velocities (where atoms cannot get ionized by the passing charged particle), the stopping power of a positively charged particle is proportional to its velocity yielding
  \begin{equation}
 \frac{dE}{dx}=-32 \pi^2 n \frac{\epsilon^{7/6} e^2a_0}{Z'} \frac{v}{v_0}, \label{dEdx}
 \end{equation}
 where $v_0$ and $a_0$ are the Bohr velocity and Bohr radius respectively and $Z'^{2/3}=\epsilon^{2/3}+Z^{2/3}$.
 The energy loss above can be trivially rewritten as 
 \begin{equation}
 \frac{dE}{dx}=-B \sqrt{E}, \label{dedx}
 \end{equation} 
where $B=32\sqrt{2} \pi^2 n \epsilon^{7/6} e^2a_0/Z'v_0\sqrt{m}$ ($m$ being the mass of the millicharged particle). The equation can be now easily integrated
from an initial value of kinetic energy $E_{\text{in}}$ (corresponding to the usual dark matter velocity of 230 $\text{km}/\text{s}$) to a final $E_f$. We can find the upper value of $E_f$ in order for a collision to produce enough recoil to trigger the detector. The recoil energy $T$ takes values between $0<T<4E_fmm_N/(m+m_N)^2$. A uniform  distribution of $T$ is usually assumed within this range. If we demand the recoil energy to be less than 1 keV (which is already below the thresholds of current experiments), we can estimate what minimum value of $E_f$ can produce this recoil. By solving the differential Eq.~(\ref{dedx}) and tracking the solution from $E_{\text{in}}$ to $E_f$ we can find what distance is needed to slow down the millicharged particle to an $E_f$ that cannot produce a detectable recoil in the underground detectors. We find that for an $\epsilon=10^{-5}$ and $m=1$ TeV, this distance is a few  meters. For this we use an average matter density of $5.5\text{g}/\text{cm}^3$.  CDMS lies 2341 feet below the surface and Xenon lies below 3700 meters water equivalent. Clearly no constraints can be imposed on both versions of dark pion: the first one because the particle is neutral with constituents of $\epsilon \sim 10^{-12}$ that is too small to produce a significant rate, and the second (with $\epsilon \sim 10^{-5}$) because the particle could slow down significantly before it reaches the detector. Assuming a depth of that of the site of Xenon, we find that a TeV millicharged particle  can slow down sufficiently to avoid direct detection for any $\epsilon$. However, two things have to be taken into account. The first one is that one should consider in addition direct collisions of the millicharged particle with the nuclei. This leads to faster loss of energy. The second point is that Eq.~(\ref{dEdx}) might not be accurate since for $\epsilon<<1$ some of the approximations leading to Eq.~(\ref{dEdx}) might not be as good as in the case where $\epsilon\sim 1$. A more thorough study of the stopping power of a slow moving millicharged particle is needed.

Finally there is another point that enables a thermally produced dark pion to evade detection. As it was pointed out in~\cite{Chuzhoy:2008zy}, for
$100\epsilon^2 \lesssim m \lesssim 10^8 \epsilon$ TeV, galactic magnetic fields prevent millicharged dark matter from the halo to enter the galactic disk and those initially trapped in the disc are accelerated and injected out of the disc within 0.1 to 1 billion years. Thus no constraints on millicharged dark matter can be imposed based on direct search experiments within the above range.  For $\epsilon\simeq 10^{-5}$ the GeV-TeV mass range of the millicharged particle is inside the above nondetection region.

\subsection{Anomalous Isotope Constraints}
In this section we present generic constraints valid for any type of millicharged dark matter particles based on searches of anomalous isotopes of hydrogen and oxygen in the sea water and helium in the atmosphere.

The binding energy between a (negatively) millicharged heavy WIMP with a nucleus is 
\begin{equation}
E_{\text{Bin}}=\frac{\epsilon^2 Z^2 \alpha^2 m_N}{2} \label{bin1}
\end{equation}
where we assumed that $m_N<<m$. Typical examples of couplings with hydrogen and oxygen nuclei give correspondingly
\begin{eqnarray}
E_H\simeq 2.5 \times 10^4 \epsilon^2 \rm eV \nonumber \\
E_O \simeq 2.6 \times 10^7 \epsilon^2 \rm eV. \label{bin}
\end{eqnarray}
The kinetic energy of a WIMP assuming a velocity $v=230 \rm km/sec$ is 
 \begin{equation}
 E_k \simeq 2.9 \times 10^2 \frac{m}{\rm GeV} \rm eV.
 \end{equation}      
  
First we can calculate the cross section for the capture of a millicharged particle by a nucleus. If the Bohr radius of a bound state between the nucleus and the milli-charged particle is smaller than the corresponding atomic one (nucleus-electron system), one can use to good approximation the formula that gives the cross section for a capture of an electron by a charged ion~\cite{Landau}
\begin{equation}
\sigma_N = \frac{2^7\pi}{3}Z^5 \alpha^3 a_0^2 \left ( \frac{E_{\text{Bin}}^0}{E} \right )^{5/2}. \label{for1}
\end{equation}
This is the cross section for the capture of an electron by a nucleus of charge $Ze$, where $\alpha$ is the fine structure constant, $E$ is the energy of the electron, and $a_0$ and $E_{\text{Bin}}^0$ are respectively the Bohr radius and the binding energy of the hydrogen atom. This formula has been derived at the limit where a moving light electron gets trapped by a heavy nucleus at rest. 
In order to be able to use this formula in our case, we have to go to the rest frame of the heavy millicharged particle. Therefore in our case of interest (i.e. the capture of a millicharged particle with charge $-\epsilon e$ by a nucleus with charge $Ze$), $Z$ has to be substituted by $Z \epsilon$ in Eq.~(\ref{for1}), $a_0$ is the Bohr radius having substituted the mass of the electron with the mass of the nucleus since the reduced mass of two unequal masses is given approximately by the smaller one which in this case is the mass of the nucleus, and $E_{\text{Bin}}^0$ is the binding energy of the system nucleus-WIMP having normalized $Z=\epsilon=1$. Due to the strong dependence of capture cross section on the energy $E$, there are two distinct cases here. In the first one, the millicharged particle loses enough energy via electromagnetic interactions with electrons (Eq.~(\ref{dEdx})) and slows down to thermal velocities. Since the millicharged particle is heavier than the nuclei of the water and air that we are going to consider, the particle might be considered at rest and therefore $E$ in Eq.~(\ref{for1}) is  the thermal energy of the atoms $3k_BT/2$. If electromagnetic interactions do not slow down sufficiently the millicharged particle, $E$ is the energy of an atom moving with the velocity of dark matter $v \simeq 230\text{km}/\text{sec}$.

There are two main constraints from anomalous isotope searches that we consider, i.e. the hydrogen and oxygen ones from ocean water and the helium ones from the atmosphere. Using Eqs.~(\ref{bin}) and (\ref{for1}) and assuming that the millicharged particle has slowed down to its thermal velocity, we find the capture cross sections for this regime
 \begin{eqnarray}
\sigma_H \simeq 1.4 \times 10^{-13} \epsilon^5 \text{cm}^2 \nonumber \\
\sigma_O \simeq 1.9 \times 10^{-8} \epsilon^5 \text{cm}^2.
\end{eqnarray}
The mean free path of a millicharged particle before it gets captured by a hydrogen (oxygen) nucleus will be given by $\lambda_{H(O)}=1/n_{H(O)}\sigma_{H(O)}$, where $n_H=N_A/(9~\text{cm}^3)$ and $n_O=N_A/(18~\text{cm}^3)$ are the number densities of hydrogen and oxygen respectively in water. The probability for capturing a millicharged particle is $p_{H(O)}=2.6~\text{km}/\lambda_{H(O)}$ if $\lambda_{H(O)}>2.6~\text{km}$, or one otherwise. As we mentioned, 2.6 km is the average ocean depth. It is easy to check that due to the fact that $\sigma_O>>\sigma_H$, most captured millicharged particles are captured by the oxygen nucleus. The ratio of the number density of the millicharged particles over the density of hydrogen in the sea water will be 
\begin{equation}
\frac{n_{\chi}}{n_H}=\frac{\rho_{\chi}}{n_H m_{\chi}}\frac{p v}{2.6 \text{km}} t_{\text{acc}}, \label{acc}
\end{equation}
where $\rho_{\chi}=0.3~\text{GeV}/\text{cm}^3$ is the local dark matter density in the earth, $p$ is the probability of capture derived above, $v=230~\text{km}/\text{sec}$ is the velocity of the millicharged particle, and $t_{\text{acc}}=3 \times 10^9 \text{years}$ is the approximate age of the oceans. If $\epsilon$ is sufficiently large, then the reactions
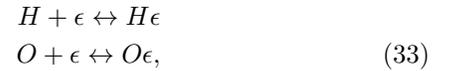
\begin{eqnarray}
H+\epsilon \leftrightarrow H\epsilon \nonumber \\
O+ \epsilon \leftrightarrow O\epsilon, \label{react}
\end{eqnarray}
are in thermal equilibrium. We take the condition for this to be $t_{\text{therm}}=1/n\sigma v<< t_{\text{acc}}$, where $v=\sqrt{3T/m}$ is the thermal velocity of a hydrogen or oxygen atom. The condition for thermalization is fulfilled for $\epsilon \gtrsim  5 \times 10^{-8}$ in the case of oxygen and $\epsilon \gtrsim 4 \times 10^{-7}$ in the case of hydrogen. Using an energy loss like the one of Eq.~(\ref{dEdx}) one can check that indeed for such a value of $\epsilon$ or higher, millicharged particles have slowed down to their thermal velocities. The reactions of Eqs.~(\ref{react}) are strictly in thermal equilibrium using the strictest of the two which is the one from hydrogen. 
 In this case we can use the Saha equations that determines the equilibrium abundances
\begin{eqnarray}
\frac{n_\epsilon n_H}{n_1}=\frac{2}{\Lambda_H^3}e^{-E_H/T} \equiv A, \nonumber \\
\frac{n_\epsilon n_O}{n_2}=\frac{2}{\Lambda_O^3}e^{-E_O/T} \equiv B, \label{saha}
\end{eqnarray}
where $n_1$ and $n_2$ are the number densities of anomalous isotopes $H\epsilon$ and $O\epsilon$ respectively, and $\Lambda_{H(O)}=\sqrt{2\pi/m_{H(O)}T}$ is the thermal De Broglie wavelength. From Eqs.~(\ref{saha}) one can solve for $n_1$ and $n_2$,
\begin{eqnarray}
n_2&=&\frac{A}{2B}n_1, \nonumber \\
n_1&=&\frac{n_{\chi}}{1+\frac{A}{2B}+\frac{A}{n_H}}, \label{n1n2}
\end{eqnarray}
where $n_{\chi}$ is the total number density of millicharged particles in the water (captured or free). There are strict constraints on $n_1/n_H$ based on hydrogen anomalous isotope searches in sea water. These constraints set an upper limit for $n_1/n_H$ at $10^{-29}-10^{-28}$~\cite{Smith}. However, it has been speculated that due to the heaviness of the millicharged particles, anomalous isotopes might concentrate at larger depths inside the oceans. Deep sea searches for hydrogen anomalous isotopes have imposed less strict upper limits of $\sim 4 \times 10^{-17}$~\cite{Yamagata:1993jq}, which are the ones we are going to use. If one uses Eqs.~(\ref{acc}), and (\ref{n1n2}), the ratio $n_1/n_H$ can be estimated and compared to the upper limit from deep water searches of anomalous hydrogen isotopes. The result is quite interesting. For a TeV particle with relatively large $\epsilon \gtrsim 1.6 \times 10^{-4}$ no constraints are imposed. Despite the relatively large values of $\epsilon$, due to the fact that the binding energy of the particle to the oxygen is much larger than that of the hydrogen (see Eq.~(\ref{bin})), the millicharged particle strongly prefers to ``hide'' in oxygen rather than hydrogen. Our estimate is in accordance with the qualitative argument presented in~\cite{Cline:2012is}.
For the rest of the $\epsilon$ phase space where the reactions~(\ref{react}) are still in thermal equilibrium i.e. $4 \times 10^{-7} \lesssim \epsilon \lesssim 1.6 \times 10^{-4}$, we find that TeV dark millicharged particles are excluded within the region $2.9 \times 10^{-6} \lesssim \epsilon \lesssim 1.6 \times 10^{-4}$.  Finally, below $\epsilon \lesssim 4 \times 10^{-7}$,
 there is no thermal equilibrium between  the reactions (\ref{react}), and therefore in that case the maximum value of $n_1/n_H$ is given directly by Eq.~(\ref{acc}) (substituting $p$ by $p_H$). In reality photons or other type of collisions could destroy part of the $H\epsilon$ bonds reducing the $n_1$ abundance. However, even with no further distraction of the bound states no constraints can be imposed in this regime.  The excluded region $2.9 \times 10^{-6} \lesssim \epsilon \lesssim 1.6 \times 10^{-4}$ for a TeV particle, expands slightly to $1.8 \times 10^{-6} \lesssim \epsilon \lesssim 1.7 \times 10^{-4}$ for a 100 GeV millicharged particle.

There are two caveats on the above derived exclusion region of $\epsilon$ which might potentially lead to narrowing or even evasion of the constraints. The first one has to do with the nature of the experimental searches for anomalous isotopes. The examined water has gone through a phase of distillation and electrolysis among others, and therefore it is possible for the tiny bond of $10^{-6}$ eV (for an $\epsilon\simeq 10^{-5}$, see Eq.~(\ref{bin})) between the millicharged particle and the hydrogen to break apart. 
  
 The second caveat has to do with the validity of Eq.~(\ref{for1}). This is in principle valid as long as the Bohr radius of the system WIMP-nucleus is smaller than that of the electron-nucleus system. If the former Bohr radius is larger than the latter, the capture cross section should be smaller, since it would be dominated by electric dipole type interactions between the millicharged particle and the system nucleus-electron.  The Bohr radius is $a=a_0m_e/(Z \epsilon m_N)$ where $m_N$ is the mass of the nucleus and $a_0$ the usual atomic Bohr radius. It turns out that for hydrogen, $a_H<a_0$ if $\epsilon > 5 \times 10^{-4}$ while for oxygen $a_O<a_0$ if $\epsilon>4.2 \times 10^{-6}$. If $a>a_0$ the capture cross section is expected to be lower that that of  Eq.~(\ref{for1}) because among other, the binding energy of the millicharged-nucleus system will become smaller.  The study of this type of dipole interactions goes beyond the scope of our paper.
 
  In our case of interest, searches for anomalous isotopes of oxygen might in principle give stricter constraints due to the fact that most of the millicharged particles will bind with oxygen in water. Although upper bounds on $n_\epsilon/n_O\simeq 10^{-16}$ exist in literature~\cite{Middleton:1979zz}, it is questionable if they can apply in our case due to the fact that the search took place within a limited range of masses and also due to the fact that the search did not take place on deep ocean waters. This is crucial since it is expected that $H_2O\epsilon$ will sink to large depths due to the heaviness of the millicharged particle. In addition, both caveats we presented for the hydrogen constraints apply also here.
   Even if one ignores the aforementioned issues about the deep water, mass range and breaking of the bond,  we find for a millicharged mass of 100 GeV (1 TeV) that for $\epsilon \lesssim 5 \times 10^{-6}$ ($\epsilon \lesssim7.9 \times 10^{-6}$) the oxygen constraints are evaded.

Finally we examine possible constraints that can emerge from searches of anomalous isotopes of helium in the atmosphere.
Searches for anomalous isotopes of helium in the air have also  imposed strict upper limits on the ratio of concentrations of anomalous helium over the normal one $r_{He}$ of $10^{-12}$ to $10^{-17}$~\cite{Mueller:2003ji}. Following our treatment of the hydrogen and oxygen constraints from water,  we examine the three basic reactions of interest
 \begin{eqnarray}
He+\epsilon \leftrightarrow He\epsilon \nonumber \\
O+ \epsilon \leftrightarrow O\epsilon \nonumber \\
N+ \epsilon \leftrightarrow N\epsilon, \label{react2}
\end{eqnarray}
since the two dominant components of atmospheric air is nitrogen, and oxygen. As before, one can estimate $t_{\text{therm}}$ to establish thermal equilibrium among the reactions (\ref{react2}). For helium (which obviously gives the strictest time scale), in order for $t_{\text{therm}}<< t_{\text{acc}}$, $\epsilon \gtrsim 5 \times 10^{-5}$. We have used an ``effective height'' for the atmosphere $h=20$ km in order to get an estimate on $n_{He}$. Helium is significantly younger in the atmosphere with $t_{\text{acc}}=2 \times 10^6\text{years}$~\cite{Mueller:2003ji}. Using an energy loss like the one of Eq.~(\ref{dEdx}) we find that for $\epsilon \gtrsim 5 \times 10^{-5}$, the millicharged particles have slowed down to their thermal velocities. 
The corresponding Saha equations are
\begin{eqnarray}
\frac{n_\epsilon n_{He}}{n_1}=\frac{2}{\Lambda_{He}^3}e^{-E_{He}/T} \equiv A, \nonumber \\
\frac{n_\epsilon n_O}{n_2}=\frac{2}{\Lambda_O^3}e^{-E_O/T} \equiv B, \nonumber \\
\frac{n_\epsilon n_N}{n_3}=\frac{2}{\Lambda_N^3}e^{-E_N/T} \equiv C,\label{saha2}
\end{eqnarray}
where $n_1$, $n_2$, and $n_3$ are respectively the densities of $He\epsilon$, $O\epsilon$, and $N\epsilon$.
The probability of capturing a millicharged particle into a bound state with an atmospheric nucleus inside the atmosphere is $p=n\sigma h$ where $n$ is the number density of the nucleus (oxygen, nitrogen or helium) in the atmosphere, $\sigma$ the cross section given by Eq.~(\ref{for1}) and $h$ is the effective atmosphere height. For example in the case of helium, $n_{He}=N_{He}/4\pi R^2 h$, where $N_{He}$ is the total number of helium atoms in the atmosphere and $R$ the radius of the earth. The total accumulated number of millicharged particles over $N_{He}$ is 
\begin{equation}
r_{He}=(\rho_{\chi}/m)4 \pi R^2 t_{\text{acc}}v p/N_{He}=(\rho_{\chi}/m)t_{\text{acc}}v \sigma. \label{helium_direct}
\end{equation}
 For $\epsilon \gtrsim 5 \times 10^{-5}$ (i.e. if the reactions (\ref{react2}) are in thermal equilibrium), one can estimate the production of anomalous helium 
\begin{equation}
n_1= \frac{n_{\chi}}{1+\gamma_2A/B +\gamma_3 A/C +A/n_{He}}, \label{NOHe}
\end{equation}
where $n_{\chi}$ is the total density of millicharged dark matter in the air. Despite oxygen representing a fraction 0.21 (instead of 0.78 for nitrogen) of the atmospheric gas, it gives the highest contribution due to the $Z$ dependence of Eq.~(\ref{for1}). The factors $\gamma_2=2 \times 0.21/(5.2 \times 10^{-6}) $ and $\gamma_3=2 \times 0.78/(5.2 \times 10^{-6})$ represent respectively the ratio of abundances of atomic oxygen over helium and atomic nitrogen over helium. The factors of 2 account for the fact that oxygen and nitrogen are in the molecular form $O_2$ and $N_2$. Helium consists only a tiny $5.2 \times 10^{-6}$ portion of the atmospheric air.  The region $ 6.8 \times 10^{-5} \lesssim \epsilon \lesssim 1.5 \times 10^{-4}$ where Eq.~(\ref{saha2}) are valid, is excluded. In this regime, the millicharged particles are captured efficiently by oxygen and nitrogen atoms and due to thermal equilibrium a significant amount is transferred to helium. As in the case of water, once $\epsilon$ becomes sufficiently large ($\epsilon \gtrsim 1.5 \times 10^{-4}$), due to the large binding energies of the millicharged particle with oxygen and nitrogen, no constraints from helium arise. For $\epsilon \lesssim 5 \times 10^{-5}$, although Eqs.~(\ref{saha2}) are not valid overall, because reactions (\ref{react2}) are not in thermal equilibrium, collisions of the $He\epsilon$ system with photons or other atoms might lead to destruction of the $He\epsilon$ population. In this case the first
equation of (\ref{saha2}) might give an upper bound for $n_1$. $n_1/n_{He}$ can be easily estimated from Eq.~(\ref{NOHe}) if one sets $\gamma_2=\gamma_3=0$, and uses $n_{\chi}/n_{He}=r_{he}$ from Eq.~(\ref{helium_direct}). Upon making the above assumption no constraints can be imposed below $\epsilon \lesssim 5 \times 10^{-5}$. Therefore the exclusion region for a TeV millicharged particle is $6.8 \times 10^{-5} \lesssim \epsilon \lesssim 1.5 \times 10^{-4}$. This expands to $5 \times 10^{-5} \lesssim \epsilon \lesssim 1.5 \times 10^{-4}$ for a 100 GeV particle.

\section{Conclusions}

We presented a model of composite dark matter with millicharged constituents. We showed under what conditions such a candidate can provide a monochromatic 130 GeV photon line via decays. We found that for a compositeness scale of order TeV, $\epsilon \sim 10^{-12}$. We also estimated that for $\epsilon \sim 10^{-5}$, millicharged dark matter can be produced thermally via the freeze-in mechanism. Due to self-interactions this dark matter candidate might be warmer than cold dark matter. We studied all possible constraints coming from self-interactions, the early Universe and BBN. We also studied possible constraints from direct search experiments. Unlike what was thought until now, we demonstrated that millicharged dark matter cannot be
excluded easily in underground detectors. For very small values of $\epsilon$ the cross section is low, but even for intermediate values where the cross section can be sufficiently large, millicharged particles might decelerate substantially providing recoil energies below the threshold of the detectors. Finally we examined possible constraints arising from searches of anomalous hydrogen in deep ocean water, oxygen and helium in the atmospheric air. We found that searches of anomalous hydrogen can potentially exclude the region (for a TeV particle) $2.9 \times 10^{-6} \lesssim \epsilon \lesssim 1.6 \times 10^{-4}$, while searches of anomalous helium
could exclude a tiny region around $\epsilon \simeq10^{-4}$. The candidate that provides the Fermi line, passes all these constraints. Millicharged dark matter thermally produced via the freeze-in mechanism seems to fall into the exclusion region of the hydrogen. However, these constraints have to be taken with caution since the experimental procedure for the detection of the anomalous atoms might destroy the weak bond of the nucleus-WIMP system. In addition TeV millicharged dark matter with $\epsilon \sim 10^{-5}$ can evade these constraints because it falls into the region which  galactic magnetic fields prevent the particles from entering the galactic disk and those initially trapped in the disc are accelerated and injected out of it relatively fast.
\section{Acknowledgements}
C.K. is supported by the Danish National Research Foundation, Grant No. DNRF90.
   
  \end{document}